\documentclass[useAMS,usenatbib,usegraphicx]{mn2e}

\title[Magnetic field in the solar convection zone]{On the magnetic field required for 
driving the observed  angular-velocity variations in the solar convection zone}
\author[Antia, Chitre and Gough]{H. M. Antia$^{1}$\thanks{E-mail:
antia@tifr.res.in}, S. M. Chitre$^{2,3,5}$ and D. O. Gough$^{3,4,5}$\\
$^{1}$Tata Institute of Fundamental Research, Homi Bhabha Road, Mumbai 400005,
India\\
$^{2}$Centre for Basic Sciences, University of Mumbai, Mumbai 400098, India\\
$^{3}$Institute of Astronomy, University of Cambridge, Madingley Road,
Cambridge CB3 0HA, UK\\
$^{4}$Department of Applied Mathematics and Theoretical 
Physics,  Wilberforce Road, Cambridge CB3 0WA, UK\\
$^{5}$Hansen Experimental Physics Laboratory, Stanford University, CA 94305, USA}
\begin{document}



\maketitle

\label{firstpage}

\begin{abstract}
{A putative temporally varying circulation-free magnetic-field configuration is inferred in an equatorial segment 
of the solar convection zone from the helioseismologically inferred  angular-velocity variation, 
assuming that the predominant dynamics is angular acceleration produced by the azimuthal Maxwell stress  
exerted by a field whose surface values are consistent with photospheric line-of-sight measurements.
}
\end{abstract}


\section{Introduction}
This paper reports a first step towards investigating directly from helioseismological inference 
of angular-velocity variations 
the magnetohydrodynamical force balance in the solar convection zone.  The balance is predominantly azimuthal, and we presume that rotational acceleration is caused by Maxwell stresses.  We ignore all other forces.

Our long-term goal is to infer from seismic and surface-magnetic-field data the 
distribution of the axially symmetric component of the magnetic field throughout the convection zone, and possibly beneath, together with the meridional circulation associated with it.  Our intention is to accomplish that task by an inverse dynamical argument, starting from what one might regard as an outcome of the dynamics, namely the temporally varying angular velocity, which has already been determined seismologically.
The task is thus to find a magnetic-field configuration and a circulation pattern that both satisfy the magnetic induction equation and produce a Maxwell stress which, together with the transport of angular momentum by both the meridional flow and putative Reynolds stresses, induce the observed angular velocity as a consistent consequence of the azimuthal component of the (angular) momentum equation.  

Buoyancy-driven motion in the convection zone redistributes angular momentum and magnetic field, the larger-scale motion -- what is normally termed rotation, zonal flow
and meridional circulation -- via laminar advection, the smaller scales via anisotropic Reynolds stresses and turbulent diffusion.  Usual (forward) theoretical investigations are aimed at reproducing that motion and its consequences, such as the temporal variation
in the angular velocity, which is then to be compared with observation.   To that end it is necessary to solve the full coupled momentum and magnetic induction equations.  Our intention here is different.  We argue inversely: to infer from the observed variation in the angular velocity what force is required to drive it.  In this first foray we assume 
axisymmetry; in that case only the azimuthal component of the momentum equation is required for inferring the pertinent components of the stress.  It is not necessary to consider the meridional components of the momentum equation;  in principle they can be used subsequently to infer from our results the forces that control the flow, and thereby provide a 
valuable diagnostic for forward dynamical calculations.  That lies outside the domain of our present endeavour, so we do not discuss it here. 
Nor do we discuss the underlying mechanisms that maintain the variation observed.

The problem thereby presented to us is highly nonlinear. Our hope is that it can be solved by iteration.  That will not be straightforward, partly because it is necessary first to obtain a trial that is close enough to the solution to render the procedure convergent. We need also to discover an appropriate iterative procedure.  To this end we first attempt to find a solution with no meridional flow.  And we go no farther than that in this preliminary exercise.  That task is also nonlinear, although apparently less so.  We have succeeded so far in solving it by iterative back-substitution in an equatorial region occupying latitudes up to $45^{\circ}$.

The primary data for our investigation are the temporal variations over the past one-and-a-half solar cycles of the angular velocity $\Omega$ in the solar interior, obtained from  rotational splitting of seismic modes produced by advection and Coriolis precession.  The
most prominent feature of the variation is a series of latitudinal bands of faster- and slower-than-average rotation, 
which match the so-called torsional oscillations observed by Howard and LaBonte (1980) in the photosphere.  
At low latitudes, the bands appear to
drift towards the equator; at high latitudes they migrate
polewards (Howe et al.~2000; Antia \& Basu 2001).  They persist throughout the 
convection zone, and possibly extend into the radiative region immediately
beneath (Vorontsov et al.~2002; Basu \& Antia 2003;
Howe et al.~2005; Antia, Basu and Chitre~2008). The pattern seems to be
well correlated with the magnetic butterfly diagram (Antia, Basu and Chitre~2008;
Sivaraman et al.~2008), and it is commonly believed to result from the interaction
between differential rotation and magnetic field (e.g., Covas et al.~2000;
Rempel 2006).   Thus the properties of zonal flows provide a strong constraint on the dynamics of the interior of the Sun.

The frequencies $\omega$ of the seismic oscillations are also affected directly by the magnetic field: 
modes of like order $n$ and degree $l$  are split with respect to azimuthal order $m$.  The frequency splitting is very small, so it would be difficult to infer the structure of the field even if it were known that it is a  magnetic field that is the cause of the splitting.  
Like the effect of any other purely structural asphericity, the splitting that is induced is an even function of $m$, because the Lorentz force cannot distinguish between east and west.  Indeed, Zweibel and Gough (1995) have argued that one can never distinguish a magnetic field from other structural asphericity from seismic frequencies alone.  Rotation, on the other hand, is axial, and does so distinguish; it 
gives rise to frequency perturbations with an odd component  --  the term first-order in $\Omega/\omega$ is purely odd in $m$, and is used to infer $\Omega(r,\theta,t)$;  once that has been  determined, the
second-order perturbations  can be calculated and subtracted from the observed
even splitting  to yield a residual contribution which is due to the remaining component of structural asphericity, and which could be 
the direct result of a magnetic field.  If one were then to assume that the resultant even contribution is due solely to magnetism it may be possible to make some direct
seismological inference about the structure of the field. 

The contribution to
frequency splitting due to a large-scale magnetic field can be calculated
using the formulation given by Gough \& Thompson (1990).  From that analysis,  Antia, Chitre \& Thompson (2000, 2003) have inferred the possibility of a
magnetic field of about 70 kG in the vicinity of  $r=0.9R_\odot$  (cf. Kosovichev, 1996), and an upper
limit of 300 kG  on a possible field near the base of the convection zone.
Chou \& Serebryanskiy (2005) used frequencies from the Michelson Doppler Imager (Scherrer et al. , 1995) to detect some temporal variation 
correlated with solar activity which could be due to magnetic-field variation near the base of the convection zone of intensity several tens kG, and 
Baldner et al.~(2009) have tried to infer magnetic-field variation in
the near-surface regions from seismic frequencies alone.  

Here we address the rotational dynamics directly.   We enquire what force is required
to induce the observed variation in the sun's angular velocity, 
under the assumption that the driving force is solely magnetic, 
an assumption that was made in early studies of the dynamics of the torsional oscillations (Sch\"ussler, 1981; Yoshimura, 1981),
and we deduce the magnetic field that produces it.  
In this tentative foray we ignore viscous and Ohmic diffusion, be they laminar or
turbulent.  We also ignore meridional advection and its contribution to the distortion 
of the magnetic field.  We recognize that this highly simplifying assumption does not take into
account all the pertinent dynamical processes that might account for the
variation, but we present it as the first step towards obtaining a dynamical
description of the solar cycle that is  consistent with observation.

An objective superficially similar to ours appears to have been 
the motivation of a recent study by Lanza (2007). Lanza does not adopt the 
helioseismologically inferred angular velocity, but instead assumes that 
the angular velocity in the convection zone results simply by pure diffusion from the angular velocity observed at the solar surface.  Therefore his analysis bears little resemblance to ours. In his computations he ignores meridional flow, as do we.  Moreover, Lanza assumes that the principal 
component of the Maxwell stress that drives the flow is $B_r B_\phi$, and he ignores $B_\theta B_\phi$;  we take both into account, and we find that in most places it is the latter that dominates.

Our procedure is to force a balance between the azimuthal 
component of the Maxwell stress and the observed angular acceleration. 
To that end, we use the seismologically inferred angular velocity 
$\Omega(r,\theta,t)$ that was used in a recent study of the temporal 
variations in the Sun's angular momentum and rotational kinetic energy 
(Antia, Chitre \& Gough 2008). In relating the Maxwell stress to the magnetic 
field, ${\bf B}$, we confine attention to low latitudes where we presume (initially) 
that the latitudinal component of the field dominates over the radial 
component. That 
permits an iterative procedure starting from a state in which 
the radial field is ignored, ending with a state in which all field components are  
consistently accounted for. At high latitudes it is the radial component 
that dominates;  in principle a corresponding procedure might also  
be applied, but there the angular velocity is less well determined and 
therefore so is the Maxwell stress; moreover, the mathematical structure of the iterative procedure  
is different.  We suspect that  convergence is less secure.

Since angular velocity is the only property of the zonal flow 
that we have observed, we must characterize the Maxwell stress with an
effective axisymmetric magnetic field.  Moreover, because only the
north-south symmetric component of $\Omega$ is available from the
inversions of seismic frequency data, we adopt a corresponding symmetry in ${\bf B}$,
which, in accord with observations of the large-scale dipolar component in and above the photoshere, 
we presume to change polarity between sunspot cycles.
In this investigation we do not attempt to satisfy the latitudinal component
of the momentum equation. Instead, we take the observed angular velocity as our starting point.  The component of the Maxwell stress that drives that flow also drives a meridional circulation,  which
advects both magnetic field and angular momentum (moderated by diffusion), thereby complicating
the situation considerably. We have not succeeded in determining a diffusion-free flow
that satisfies the consequent dynamical balance; we recognize that one may not even exist.  One can easily calculate a
flow that arises from the field we have calculated by ignoring advection and diffusion,
but subsequent iteration to accommodate advection based on straightforward back-substitution, 
perhaps unsurprisingly,  does not converge.
To make further progress therefore demands a more sophisticated approach.

We relate the Maxwell stress to the latitudinal and azimuthal components
of the magnetic field ${\bf B}$ via the corresponding components of the
magnetic induction equation from which, subject to our assumptions, the
horizontal component of ${\bf B}$ is determined in terms of an integral of
$\Omega$. The radial component of ${\bf B}$ is then determined from the
divergence equation $\nabla\cdot{\bf B}=0$. Completing the solution
requires the imposition of boundary conditions; we use the azimuthally averaged 
observed line-of-sight
component of the magnetic field in the solar atmosphere.

\section{Formulation}
We ignore the meridional flow velocity together with perturbations to the
hydrostatic stratification induced by that flow and by the angular velocity, and indeed 
also the magnetic field itself.  
Thus, with respect to spherical polar co-ordinates $(r,\theta,\phi)$ referred to the axis of rotation,  the
azimuthal component of the diffusion-free momentum equation 
for an axisymmetric configuration is
\begin{equation}
4\pi r \rho \frac{\partial v_\phi}{\partial t} =
\frac{1}{\sin^2\theta}\frac{\partial}{\partial\theta}(\Gamma\sin^2\theta)
+\frac{1}{r^2}\frac{\partial}{\partial r}(r^3\Delta)=: 4 \pi r L_\phi\,,
\label{eq:mom}
\end{equation}
where $t$ is time, $v_\phi=r\Omega\sin\theta$ is the velocity of rotation,
and
\begin{equation}
\Gamma=B_\theta B_\phi,\qquad \Delta=B_r B_\phi,
\label{eq:gam}
\end{equation}
in which ${\bf B}=(B_r,B_\theta,B_\phi)$.  The corresponding component of
the diffusion-free magnetic induction equation is
 
 \begin{equation}
\frac{1}{\sin\theta}\frac{\partial B_\phi}{\partial t}=
B_\theta\frac{\partial\Omega}{\partial\theta}+rB_r\frac{\partial\Omega}
{\partial r}\,,
\label{eq:azi}
\end{equation}
 whence
\begin{equation}
\frac{1}{2\sin\theta}\frac{\partial B_\phi^2}{\partial t}=
\Gamma\frac{\partial\Omega}{\partial\theta}+r\Delta\frac{\partial\Omega}
{\partial r}\,.
\label{eq:ind}
\end{equation}
We confine attention to latitudes equatorward of $45^{\circ}$, where we expect  $|B_r|\ll|B_\theta|$,
implying $|\Delta|\ll|\Gamma|$. We assume also that the meridional field
has the same symmetry as a dipole: $B_r(r,\pi-\theta,t)=-B_r(r,\theta,t)$
and $B_\theta(r,\pi-\theta,t)=B_\theta(r,\theta,t)$, implying that the
azimuthal component, which is generated by stretching due to rotational shear
with $\Omega(r,\pi-\theta,t)=\Omega(r,\theta,t)$, satisfies
$B_\phi(r,\theta,t)=-B_\phi(r,\pi-\theta,t)$, whence $B_\phi(r,\pi/2,t)=0$.
Under these symmetries, equation (\ref{eq:mom}) can be integrated with
respect to $\theta$ to yield
\begin{equation}
\Gamma=-\frac{1}{\sin^2\theta}\int_\theta^{\pi/2} \sin^2\theta'\left[
4\pi\rho r\frac{\partial v_\phi}{\partial t}-\frac{1}{r^2}
\frac{\partial}{\partial r}(r^3\Delta)\right]\,{\rm d}\theta'\,,
\label{eq:t1}
\end{equation}
which is then combined with equation (\ref{eq:ind}) and integrated to
obtain $B_\phi^2$. In carrying out that procedure we notice that
$\Omega(r,\theta,t)$ may be written as $\Omega_0(r,\theta)+\Omega_1(r,\theta,t)$
with $|\Omega_1|\ll|\Omega_0|$. Accordingly, it is expedient to use this
separation for $\partial\Omega/\partial\theta$ to obtain
\begin{eqnarray}
\lefteqn{B_\phi^2=-\frac{8\pi\rho r^2}{\sin\theta}\left\{ \frac{\partial\Omega_0}
{\partial\theta}\int_\theta^{\pi/2}\Omega_1(r,\theta',t)\sin^3\theta'\,{\rm d}\theta'\right.}
\nonumber\\
&&\left.\phantom{\int_\theta^\pi}\qquad+\Lambda(r,\theta,t)+H(r,\theta)\right\},
\label{eq:bphi}\\
&=:& I(r,\theta,t)-\frac{8\pi\rho r^2}{\sin\theta} H(r,\theta),
\label{eq:bphih}
\end{eqnarray}
where
\begin{eqnarray}
\lefteqn{\Lambda=-\int_{t_0}^t\left\{\frac{\Delta}{4\pi\rho r}\frac{\partial\Omega_0}
{\partial r}\sin^2\theta+\right.}
\nonumber\\
&&\quad\left.\frac{1}{4\pi\rho r^4}\frac{\partial\Omega_0}{\partial\theta}
\int_\theta^{\pi/2}\sin^2\theta'
\frac{\partial}{\partial r}(r^3\Delta)\;d\theta'\right\}\,{\rm d}t',
\label{eq:lamda}
\end{eqnarray}
in which $t_0$ is some fiducial origin of time; $H(r,\theta)$ is a function
of integration, depending, of course, on the choice of $t_0$. For $t_0$
we use the instant at which seismic data are available.
In equations (6) and (8) we have neglected the temporally varying terms in
$\partial\Omega/\partial\theta$ and $\partial\Omega/\partial r$   
to emphasize from where the major time dependence in the solution arises; the approximation modifies
coefficients multiplying relatively small terms by no more than 0.3\%, which makes an 
imperceptible difference to the solution at the resolution of Figures 1--3.   Given $B_\phi$, the other horizontal component of the field can be obtained
as $\Gamma/B_\phi$ from equation (\ref{eq:t1}), and then $B_r$ can be obtained
by integrating $\nabla\cdot{\bf B}=0$. We point out that one might have
expected an additional function of integration, namely $M(r,t)$, added to the
right-hand side of equation (\ref{eq:t1}); it is straightforward to show
that this function leads to a component of the magnetic field that
does not satisfy the required symmetries, and must therefore
be null.

We anticipate that at any given location $B_\phi$ changes sign at some
instant during the solar cycle. That can occur only when $B_\phi^2=0$.
Therefore, we choose the function $H$ of integration to render it 
possible: 
\begin{equation}
H(r, \theta)=\frac{\sin\theta}{8\pi\rho r^2}\inf_t[I(r,\theta,t)]\,.
\label{eq:hfun}
\end{equation}
The sign of $B_\phi$ is chosen to match the sign of the observed field
near the surface;  
$B_\theta$ changes sign at other instants at which $\Gamma$ vanishes.

We introduce a flux function, $\chi$, which satisfies
\begin{equation}
(B_r,B_\theta)=\frac{1}{r\sin\theta}\left( \frac{1}{r}\frac{\partial\chi}
{\partial\theta},-\frac{\partial\chi}{\partial r}\right),
\label{eq:chi}
\end{equation}
which can be integrated to
\begin{equation}
\chi(r,\theta,t)=\sin\theta\int_r^R r'B_\theta\,{\rm d}r'+\chi(R,\theta,t)\,,
\label{eq:chif}
\end{equation}
in which the second term is determined from the longitudinally averaged
line-of-sight magnetic field $\overline{B}_\parallel(\theta,t)$ observed at the surface
$r=R$ of the Sun: 
\begin{equation}
\chi(R,\theta,t)=R^2\int_\theta^{\pi/2}(B_\theta\cos\theta' - \overline{B}_\parallel)\,{\rm d}\theta'\,.
\label{eq:chir}
\end{equation}

These equations have been solved by iterative back-substitution: expressions (\ref{eq:t1})--(\ref{eq:bphih}) and (\ref{eq:hfun}) for $\Gamma$ and $B_\phi^2$ were
evaluated first with $\Delta=0$ and $\Lambda=0$, from which a trial $B_\theta$, and subsequently 
$\chi$ and $B_r$, were obtained from equations (\ref{eq:gam}),  (\ref{eq:chif}) and (\ref{eq:chir}), and (\ref{eq:chi}).  New values of
$\Delta$ and $\Lambda$ were then evaluated,  
and the whole procedure repeated till the solution converged.

\section{Seismic data and surface magnetic field}
We have used GONG frequency-splitting data (Hill et al.~1996) collected over a period from 1995  to 2011 (GONG months 1--165).  Each data set covers three overlapping GONG months (of  36 consecutive days each), with midpoints separated by 1 GONG month.   
From each set the angular velocity $\Omega$ was inferred as a function of 
radius and latitude by  2-d RLS data fitting, implemented in the manner of  
Antia, Basu and Chitre (1998). The results were then smoothed at each
radius and latitude by fitting them, by least squares, to the oscillatory form:
\begin{equation}
\Omega(r,\theta,t)=\langle\Omega\rangle +\sum_{k=1}^2
a_k \sin(k\omega_0 t+\phi_k)=:\Omega_0+\delta\Omega\,,
\end{equation}
in which the angular brackets denote
average over time,  and  $\omega_0$ is the frequency corresponding to a period of 11.7y, 
the inferred length of solar cycle 23 (Antia \& Basu 2010);  thus $\langle\Omega\rangle$ represents $\Omega_0$ and $\delta\Omega$ represents $\Omega_1$ in equations (6) and (8).  The temporal expansion was taken to second order in order to 
accommodate quadratic nonlinearities.  Note that $\langle\Omega\rangle$, 
$a_k$ and $\phi_k$ are functions of $r$ and $\theta$.  The smoothed angular velocity and its temporal derivatives
were used to deduce the temporally varying magnetic field using the procedure described 
in \S2.
 
Full-disc magnetograms from the Wilcox Solar Observatory have long provided valuable data for global magnetic studies (e.g. Duvall et al., 1979).  Here we use data obtained by Lo, Hoeksema and Scherrer (2010), kindly provided by T. Hoeksema in the form of a longitudinally averaged line-of-sight magnetic field $\overline{B}_\parallel(\theta,t)$ -- the zeroth moment of the magnetograms with respect to longitude -- which constitutes our surface boundary condition.

\section{Results}

The starting point of our analysis is the observed temporally varying component $\delta \Omega(r,\theta,t)$ of the angular velocity.  It is depicted in Figure 1 in the equatorial sector of a meridional quadrant  within which our analysis has been carried out, at epochs separated, in the main,  by two years.  At any instant there are typically three principal regions, with respect to latitude, of coherent variation; they are of alternating sense, and they extend over the entire vertical range of the convection zone, although the boundaries are not vertical: the pattern gradually distorts.  However, the predominant variation is a migration towards the equator at a rate which causes at (almost) 
every fixed location an oscillation with the period of the sunspot cycle, as has been observed at the the surface of the Sun (Howard and LaBonte, 1980) and in the subphotospheric layers of the convection zone (e.g. Vorontsov et al.~2002).  The sign of $\delta \Omega$ changes rather later than the sunspot maximum and sunspot minima. 
The maximum amplitude of $\delta \Omega$ is a little above 1.2 nHz.
Note that the Lorentz force density, $L_\phi$, also shows temporal
variation, and has typically two main regions in latitude extending over the
entire vertical range of the convection zone.  Its maximum magnitude 
occurs around $0.8 {\rm R}_\odot$. 

The dominant Maxwell stress that drives the flow is proportional to $\Gamma = B_\theta B_\phi$ (see equation (1)).  It is displayed in Figure 2,  alongside the other pertinent component, $\Delta$.   There is little obvious large-scale migration in 
the convection zone,  unlike the variation of $\delta \Omega$ from which $\Gamma$ was derived; rather, simply an alternation in sign, occurring a little before the sign change in $\delta \Omega$.  
However, the sign change does actually begin as an equatorward  migration of a tongue of oppositely signed stress centred at $r/R \simeq 0.8$, emerging into our region of consideration  from the highest latitude in 1996, 2002 and again in 2008  
(a panel displaying the last of these is included in Figure 2; it is evident in Figure 3 
that it stems principally from a similar configuration of $B_\phi$), 
and which rapidly expands  to occupy much of the convection zone.  During this process, the stress in the convection zone typically suffers two sign changes with 
respect to radius, particularly at the higher latitudes.  Note that  the contours of $\Delta=B_r B_\phi$ are substantially more 
widely separated than those of  $\Gamma=B_\theta B_\phi$, indicating that on the whole the radial component of the meridional field has a lesser influence than the 
horizontal component, as we anticipated.
The stress distribution in the tachocline beneath the convection zone is more complicated, possibly because advection of the magnetic field by a meridional circulation -- not incorporated here -- is important there;  broadly speaking, the stress is in the opposite sense to that in the convection zone immediately above, as is to be expected.  

Contours of constant flux function, $\chi$, which map the poloidal magnetic field, and contours on constant $B_\phi$, are depicted in Figure 3.  The poloidal field is predominantly dipolar near sunspot maximum, around 2000, and predominantly quadrupolar, with higher multipoles also evident, near sunspot minimum, both near 1996 and throughout the latest broad minimum from about 2005 to the present.  The dipolar component reverses  sign during the declining phase of the cycle, somewhat earlier than sunspot minimum; it appears to have a characteristic period twice that of the sunspot cycle, approximately 22~y, as is observed to be the case at the surface.  The azimuthal field reverses sign throughout most of the convection zone near sunspot minimum, and it too has a dominant period twice that of the sunspot cycle.  Its maximum intensity, near the 2001 sunspot maximum, is about 4.5~kG; today it has attained the same value, and is still rising as the next maximum is approached. The field in the tachocline is more complicated:   only near sunspot minimum is there a significant radial field, which then appears to be wound up by the shear to generate azimuthal field;  near sunspot maximum the tachocline field is predominantly horizontal. 

The temporal variation of the field is resolved better in Figure 4.  There $B_\theta$ and  $B_\phi$ are plotted against latitude and  time at various radii.  Perhaps the most striking feature is the 
upward migration  of the pattern of $B_\theta$, which is most easily seen in the location and shapes of the null lines: the time lag behind the variation at the base of the convection zone is greatest near the equator, reaching a value of about 14y at $r/R=0.98$;   
the lag is least at latitudes $\pm 30^{\circ}$, the latitudes at which sunspots first appear at 
the surface.   In addition there is a temporal overtone in $B_\theta$ near the equator, which appears to be related to the octupolar component of the poloidal field  (cf. Figure 2).  There is also an evident temporal overtone at mid latitudes near the base of the convection zone. The azimuthal component $B_\phi$ of the field also migrates upwards, although only above $r/R \simeq 0.8$; the greater 
migration speed near latitudes $\pm 30^{\circ}$ is perhaps somewhat more obvious than it is for $B_\theta$.  The azimuthal component of the field changes sign near the surface around 2006, near the beginning of the prolonged period of low activity from 2005 until about 2011.  The reversal of $B_\theta$ near the surface also starts near the beginning of the prolonged sunspot minimum, at 
latitudes near $\pm 30^{\circ}$, but occurs later at low latitudes, extending throughout the  period of low activity.  The greatest intensity, $B_{\theta, {\rm max}}$, of $B_\theta$, was about 100G,  and occurred at the beginning of the period of investigation near the the base of the convection zone; judging from the upward migration rate of the field, it is likely that the actual maximum had occurred about 2y before that.  
The radial component, $B_r$, is considerably less intense, so $B_{\theta, {\rm max}}$ estimates the maximum intensity of the total poloidal field.  The intensity of the poloidal field  appears to have attained a second maximum in 2011, with a value of only about 55G.

\begin{figure}
\includegraphics[width=90mm]{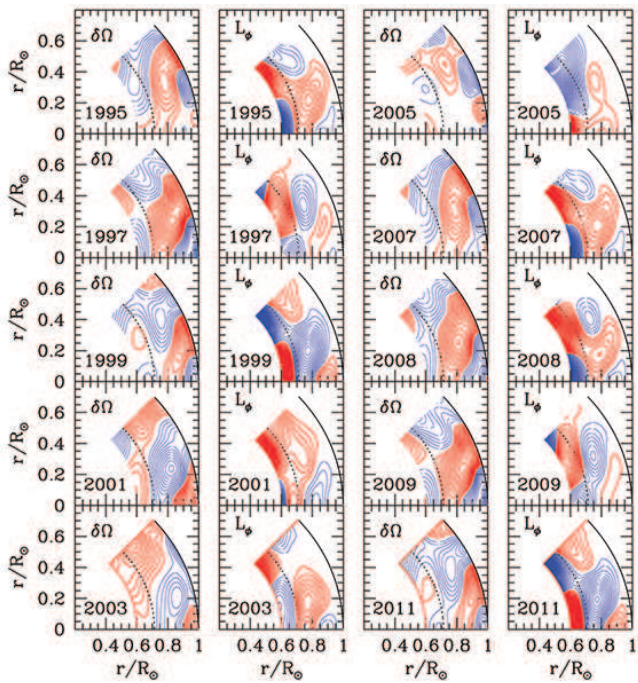}
\caption{Contours of  the angular-velocity variation $\delta \Omega$ and the Lorentz force density $L_\phi$ (defined by equation (1)\,), each averaged over three GONG months in the middles of the years indicated in the panels.  Red contours are positive (i.e. they correspond to azimuthal motion and force density $L_\phi$ in the direction of the mean rotational motion), blue negative.  The continuous black circular segment marks the photosphere, the dashed segment the base of the convection zone.  The contour spacings are 0.1\,nHz in $\delta \Omega/2 \pi$, and $5\times10^{-8}\,{\rm g}\,{\rm cm}^{-2}\,{\rm s}^{-2}$ in $L_\phi$, except that the null contours are not drawn.}
\label{fig:1}
\end{figure}

\begin{figure}
\includegraphics[width=90mm]{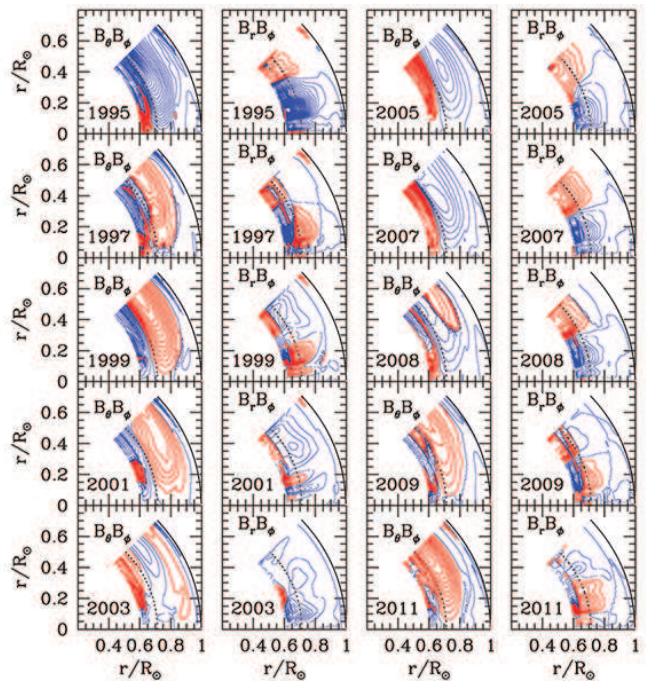}
\caption{Contours of  the components $\Gamma = B_\theta B_\phi$ and $\Delta=B_r B_\phi$ of the Maxwell stress, each inferred over three GONG months in the middles of the years indicated.  Red contours are positive, blue negative.   The contour spacings are  $(100\, {\rm G})^2$ in both, except that the null contours are not drawn. }
\label{fig:2}
\end{figure}

\begin{figure}
\includegraphics[width=90mm]{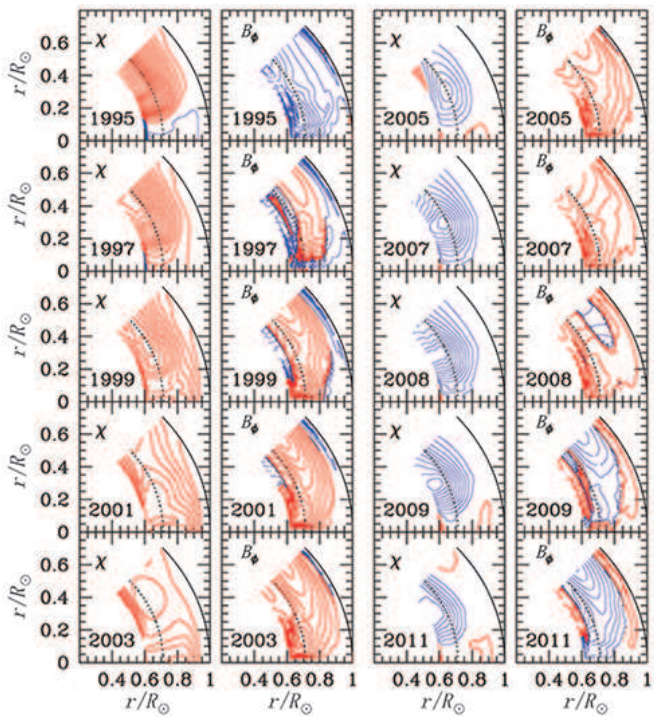}
\caption{Contours of constant flux function $\chi$, which indicate the direction of the poloidal component of the magnetic field, and the azimuthal component $B_\phi$ of the field, using the same conventions as in Figure 1:  the direction of the poloidal field is clockwise on red contours of $\chi$, anticlockwise on blue.  The contour spacing is $0.5\, {\rm G}\, {\rm R}_\odot^2$ in $\chi$ and  500\,G in $B_\phi$, except that the null contours are not drawn.}
\label{fig:3}
\end{figure}

\begin{figure}
\includegraphics[width=85mm]{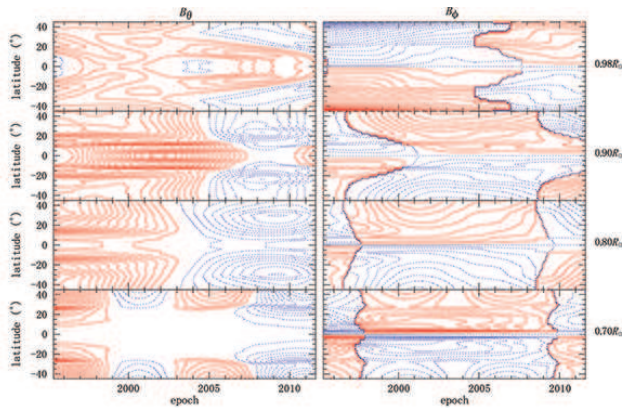}
\caption{In the left-hand panels are displayed  contours of constant $B_\theta$ inferred from three-month averages of $\delta \Omega(r,\theta,t)$ obtained by inverting  all the (overlapping)  GONG  data sets and fitting to expression (13), plotted against epoch and latitude at the four radii indicated; in the right-hand panels are displayed $B_\phi$.  
 The contour separations are 1\,G, \ 2.5\,G,  5\,G and 5\,G in $B_\theta$, and 100\,G, 250\,G, 500\,G and 500\,G in $B_\phi$ at $r=0.98,\; 0.90,\; 0.80 \; {\rm{and}}\; 0.70\,{\rm{R}_\odot}$, respectively; red represents positive values, blue negative.  
The plots 
in the southern hemisphere were obtained from the assumed symmetry relations quoted in the text. 
As in the previous two figures, the null contours are not drawn.}
\label{fig:4}
\end{figure}

\section{Discussion}

We have attempted to infer the magnetic-field configuration and its temporal
variation in the solar convection zone by dynamical reasoning, starting from the
seismically inferred angular velocity. Our analysis is based on the assumption that the Maxwell stress is wholly responsible for driving the time-varying axisymmetric component of the migrating bands of zonal flow.   We have ignored  viscous and Ohmic diffusion; we have
also ignored meridional flow.

It should be appreciated that the magnetic field in the Sun may be inhomogeneous on  small scales.    Therefore the field  intensities that we have quoted are essentially rms values characteristic of the spatially averaged Maxwell stress.

Some features of our findings are in accord with common expectation: an oscillatory, predominantly dipolar (at times of sunspot maximum), poloidal magnetic field with a 22-y period, 
which reverses near sunspot minimum, revealing a predominantly quadrupolar residual component near reversal; and a large-scale toroidal field, presumably generated by the stretching of the poloidal field by the differential rotation, which also has a period of 22y.  What is new is the 
seismological evidence that the reversal of the magnetic field migrates upwards from the base of the convection zone. 

The variation of the field that we find in the tachocline is more complicated.  
Comparison of the radial and azimuthal components of the field suggests that there is a significant meridional flow in the tachocline which oscillates with the 
cycle and which advects the field:  equatorward during the rising phase of the cycle, poleward during the declining phase.  However, we must emphasize that a 
fully consistent calculation incorporating advection has not been carried out, 
so the hint must be taken with a pinch of salt.

Our omission of meridional circulation is a serious deficiency in our calculations. 
We have assumed that the predominant force causing the rotational acceleration 
arises from Maxwell stresses.  But if a major component of the angular-momentum 
variation were to be due to advection, our results would prove to be inadequate.  Is that actually the situation?  Our failure to obtain convergence by straightforward 
iterative back-substitution, starting from the circulation-free solution 
described in this paper, and obtaining an estimate of the circulation at each iteration from the magnetic field configuration via the meridional components of 
the induction equation, suggests that it may be.  But studies of simplified, 
yet dynamically consistent,  dynamo models by Rempel (2006) suggest that the influence of circulation on the overall structure of the field may not be  
severe.  That supports our optimism that a relatively straightforward dynamical inversion procedure might be within our grasp.   What might be the necessary ingredients?

First we intend to extend our circulation-free solution to the poles.  Although 
we have found that it is possible to extend the procedure described in this paper to somewhat higher latitudes, it cannot be extended far, because the assumed 
low $B_r$:$B_\theta$ ratio adopted to justify the neglect of $B_r$ in computing our initial estimate of the field in the equatorial regions is violated near the poles.  Instead, we should 
design a procedure based on an initial state obtained by ignoring $B_\theta$;  that leads to a problem with a different mathematical structure.  It is interesting to 
observe that in completing the computation of the field structure in that case it is the first moment with respect to longitude of the the observed 
line-of-sight field that is needed for providing the magnetic boundary condition at the surface.  That is mathematically -- although hardly physically -- independent of the 
zeroth moment used in the equatorially concentrated analysis presented here.  
Therefore, the degree to which the polar and equatorial solutions match at 
mid-latitudes will provide some measure of the degree to which our underlying assumptions are valid.

Next we shall attempt to incorporate the meridional flow ${\bf v}$.  That is to be accomplished by requiring that it advects, against Ohmic diffusion, the magnetic 
field in such a manner as to yield the field ${\bf B}$ whose azimuthal stress, together 
with angular-momentum advection by ${\bf v}$, maintains the the observed angular-velocity variation $\delta \Omega$.  It is determined by the $\theta$ component of the 
 induction equation coupled with the continuity equation  $\nabla.(\rho {\bf v}) = 0$.
 To complete the integration requires an additional boundary condition: namely, the value of the product $B_r v_\theta$ at the surface, which can be obtained with the help 
of either the seismologically inferred meridional flow near the surface (and extrapolated to the surface) or observations of small photospheric tracers.

If eventually we succeed in solving the full problem, what observable predictions might there be for testing the outcome?  We cannot use the meridional flow observed in the photospheric layers because that will be needed as a boundary condition for completing the integration for the stream function.  However, the flow inferred well below the photosphere could be compared with direct helioseismological inferences, which are likely to have been extended to substantially greater depths than have been achieved to date.  In particular, it would be  interesting to see whether we reproduce the double-cell structure suggested by recent time-distance analysis (J. Zhou, personal communication, 2012).  It would be interesting also to see whether the inferred flow matches the upwelling from the tachocline predicted by Spiegel and Zahn (1992) and Gough and McIntyre (1998), although that component of the flow might be too weak to have a noticeable impact on the flow geometry above.  In any case, our eventual results would provide a serious template against which to compare dynamical theories of the convection zone.

Finally we recall that we have assumed axisymmetry on the large scale.  To study 
asymmetric deviations would require seismic inference of the nonaxisymmetric zonal flow.  That is not presently available.

\section*{Acknowledgments}

This work  used data obtained by the Global Oscillation
Network Group (GONG) project, managed by the National Solar Observatory,
which is
operated by AURA, Inc. under a cooperative agreement with the
National Science Foundation. The data were acquired by instruments
operated by the Big Bear Solar Observatory, High Altitude Observatory,
Learmonth Solar Observatory, Udaipur Solar Observatory, Instituto de
Astrofisico de Canarias, and Cerro Tololo Inter-American Observatory.
We are very grateful to Todd Hoeksema for supplying magnetic-field data 
from the Wilcox Solar Observatory, and to Tom Duvall for valuable comments. 
SMC thanks the Institute of Astronomy for supporting his visits to Cambridge,  
and SMC and DOG are grateful to P.H. Scherrer for his hospitality at the Hansen 
Experimental Physics Laboratory and for support from HMI NASA contract NAS5-02139.  DOG is thankful to the
Leverhulme Trust for an Emeritus Fellowship.  We are all grateful to Amanda 
Smith for her help in preparing the diagrams.  We thank the referee for helpful comments.

\label{lastpage}

\end{document}